\documentclass[aps,prapplied,reprint,superscriptaddress,floatfix]{revtex4-2} % Appropriate for PRApplied
\usepackage[T1]{fontenc}
\usepackage{times}
\usepackage{CJKutf8}
\usepackage{graphicx}
\usepackage{booktabs}

\usepackage{amsfonts, amsmath,amssymb, bm, graphicx}
\usepackage{siunitx}
\DeclareSIUnit{\belmilliwatt}{Bm}
\DeclareSIUnit{\dBm}{\deci\belmilliwatt}
\usepackage{amssymb}
\usepackage[english]{babel}
\usepackage{lipsum}

\newcommand{\figref}[2]{\hyperref[#1]{\ref{#1}(#2)}}
\newcommand{\figrefsub}[3]{\hyperref[#1]{\ref{#1}(#2)#3}}

\usepackage{physics}
\usepackage{lipsum}
\usepackage{changes}

\usepackage[colorlinks=true,allcolors=blue]{hyperref}
\usepackage{footmisc}

\usepackage{upgreek}

\usepackage{soul}

\makeatletter
\let\ORIbbl@fixname\bbl@fixname
\def\bbl@fixname#1{%
  \@ifundefined{languagealias@\expandafter\string#1}
    {\ORIbbl@fixname#1}
    {\edef\languagename{\@nameuse{languagealias@#1}}}%
}
\newcommand{\definelanguagealias}[2]{%
  \@namedef{languagealias@#1}{#2}%
}
\makeatother

\definelanguagealias{en}{english}

\begin{document}

\title{Predicting the future with magnons}

\author{Zeling Xiong (\begin{CJK*}{UTF8}{gbsn}熊则灵
\end{CJK*})}
\affiliation{Helmholtz-Zentrum Dresden--Rossendorf, Institut f\"ur Ionenstrahlphysik und Materialforschung, D-01328 Dresden, Germany}
\affiliation{Fakultät Physik, Technische Universität Dresden, D-01062 Dresden, Germany}

\author{Christopher Heins}
\affiliation{Helmholtz-Zentrum Dresden--Rossendorf, Institut f\"ur Ionenstrahlphysik und Materialforschung, D-01328 Dresden, Germany}
\affiliation{Fakultät Physik, Technische Universität Dresden, D-01062 Dresden, Germany}

\author{Thibaut Devolder}
\affiliation{Centre de Nanosciences et de Nanotechnologies, CNRS, Université Paris-Saclay, Palaiseau, 91120, France.}

\author{Fabian Kammerbauer}
\affiliation{Institute of Physics, Johannes Gutenberg University Mainz, 55099 Mainz, Germany}

\author{Mathias Kl\"aui}
\affiliation{Institute of Physics, Johannes Gutenberg University Mainz, 55099 Mainz, Germany}

\author{J\"urgen Fassbender}
\affiliation{Helmholtz-Zentrum Dresden--Rossendorf, Institut f\"ur Ionenstrahlphysik und Materialforschung, D-01328 Dresden, Germany}
\affiliation{Fakultät Physik, Technische Universität Dresden, D-01062 Dresden, Germany}

\author{Helmut Schultheiss}\email{h.schultheiss@hzdr.de}
\affiliation{Helmholtz-Zentrum Dresden--Rossendorf, Institut f\"ur Ionenstrahlphysik und Materialforschung, D-01328 Dresden, Germany}

\author{Katrin Schultheiss}\email{k.schultheiss@hzdr.de}
\affiliation{Helmholtz-Zentrum Dresden--Rossendorf, Institut f\"ur Ionenstrahlphysik und Materialforschung, D-01328 Dresden, Germany}

\date{\today}

\begin{abstract}
Forecasting complex, chaotic signals is a central challenge across science and technology, with implications ranging from secure communications to climate modeling. Here we demonstrate that magnons, the collective spin excitations in magnetically ordered materials, can serve as an efficient physical reservoir for predicting such dynamics. Using a magnetic vortex-state micro-disk as a magnon-scattering reservoir, we show that intrinsic nonlinear interactions transform a simple microwave input into a high-dimensional spectral output suitable for time series predictions. Trained on the Mackey-Glass benchmark, which generates a cyclic yet aperiodic time series, the system achieves accurate and reliable predictions that rival state-of-the-art physical reservoirs. We further identify key design principles: spectral resolution governs the trade-off between dimensionality and accuracy, while combining multiple device geometries systematically improves performance. These results establish magnonics as a promising platform for unconventional computing, offering a path toward scalable and CMOS-compatible hardware for real-time prediction tasks.
\end{abstract}

\maketitle

\section{Introduction}

The ability to predict chaotic time series is a longstanding challenge at the crossroads of nonlinear dynamics, machine learning, and physical computing. Chaotic systems, while governed by deterministic rules, exhibit extreme sensitivity to initial conditions, rendering long-term trajectories unpredictable. This combination of determinism and apparent randomness makes them ideal benchmarks for testing new computational architectures aimed at real-time forecasting in complex, high-dimensional systems.
Reservoir computing has emerged as a particularly promising approach, offering efficient training and natural suitability for temporal tasks~\cite{10.1162/089976602760407955,appeltantInformationProcessingUsing2011,christensen2022RoadmapNeuromorphic2022}. In this framework, a fixed dynamical system projects inputs into a high-dimensional state space, from which a simple linear readout extracts the relevant information. Crucially, this separation of nonlinear dynamics and trainable output enables hardware realizations, where intrinsic material properties can be harnessed directly. Physical implementations have already been demonstrated in diverse platforms, from optical fibers~\cite{duportAllopticalReservoirComputing2012,vandersandeAdvancesPhotonicReservoir2017,chemboMachineLearningBased2020,yanEmergingOpportunitiesChallenges2024} and memristive circuits~\cite{duReservoirComputingUsing2017,moonTemporalDataClassification2019,zhongDynamicMemristorbasedReservoir2021,zhongMemristorbasedAnalogueReservoir2022} to spintronic devices~\cite{kanaoReservoirComputingSpinTorque2019,tsunegiPhysicalReservoirComputing2019,markovicReservoirComputingFrequency2019,yamaguchi2020step,taniguchiSpintronicReservoirComputing2022,akashiCoupledSpintronicsNeuromorphic2022}, nanomagnetic arrays~\cite{gartsideReconfigurableTrainingReservoir2022,allwoodPerspectivePhysicalReservoir2023,jensenClockedDynamicsArtificial2024}, skyrmions~\cite{zazvorkaThermalSkyrmionDiffusion2019,pinnaReservoirComputingRandom2020,sunExperimentalDemonstrationSkyrmionenhanced2023,leePerspectiveUnconventionalComputing2023,benekeGestureRecognitionBrownian2024}, and magnonic systems~\cite{nakaneReservoirComputingSpin2018,pappCharacterizationNonlinearSpinwave2021,leeReservoirComputingSpin2022,nakanePerformanceEnhancementSpinWaveBased2023,namikiExperimentalDemonstrationHighPerformance2023,Nagase2024,namikiFastPhysicalReservoir2024}. Each of these leverages unique mechanisms of nonlinearity and memory.

%%%%%%%%%%%%%%%%%%%%%%%%%%%% 
\begin{figure}[]
    \centering
    \includegraphics{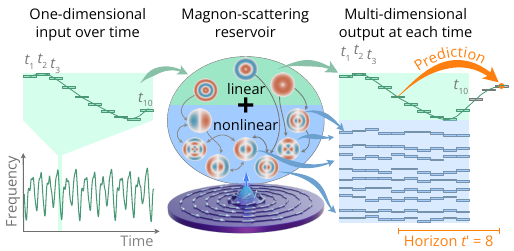}
    \caption{Principle of chaotic time-series prediction based on a magnon-scattering reservoir: a one-dimensional temporal input is encoded in a  microwave current and applied to a magnon-scattering reservoir in form of a Ni$_{81}$Fe$_{19}$ microdisk in the vortex state. Nonlinear interactions between various magnon modes transform the one-dimensional input into a multi-dimensional spectral output trained to forecast the future trajectory of the chaotic signal. }  
    \label{fig:principle}
\end{figure}
%%%%%%%%%%%%%%%%%%%%%%%%%%%% 

%%%%%%%%%%%%%%%%%%%%%%%%%%%% 
\begin{figure}[]
    \centering
    \includegraphics{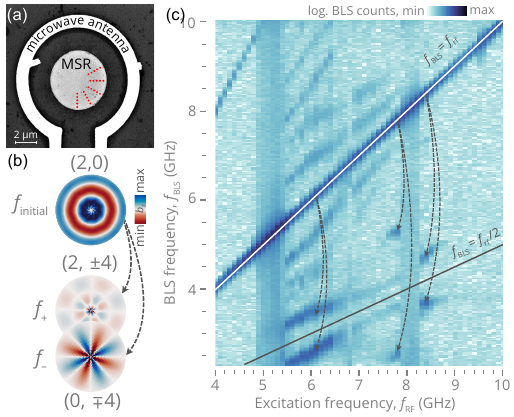}
    \caption{ (a) Scanning electron microscopy image of the magnon-scattering reservoir (MSR) in form of a 5\,\si{\micro\meter}-wide and 50\,\si{\nano\meter}-thick disk embedded in the $\Omega$-shaped microwave antenna. Red dots indicate the measurement positions used for spatial averaging of the BLS signal. (b) Schematic of a three-magnon splitting process: a directly excited mode ($2,0$) at frequency $f_\text{initial}$ decays into two secondary modes ($2,\pm 4$) and ($0,\mp 4$) with frequencies $f_{+}$ and $f_{-}$, conserving both energy and momentum.
    (c) BLS spectra measured on the device in (b) at an excitation power of 23\,dBm, i.e. above the threshold for nonlinear three-magnon splitting. Each column corresponds to a spectrum recorded at a fixed excitation frequency $f_\text{RF}$, with intensity color coded on a logarithmic scale. The strongest nonlinear response is observed for excitation frequencies between 5.5 and 8.8\,GHz.}
    \label{fig:3ms}
\end{figure}
%%%%%%%%%%%%%%%%%%%%%%%%%%%% 

%%%%%%%%%%%%%%%%%%%%%%%%%%%%
\begin{figure*}[]
    \centering
    \includegraphics{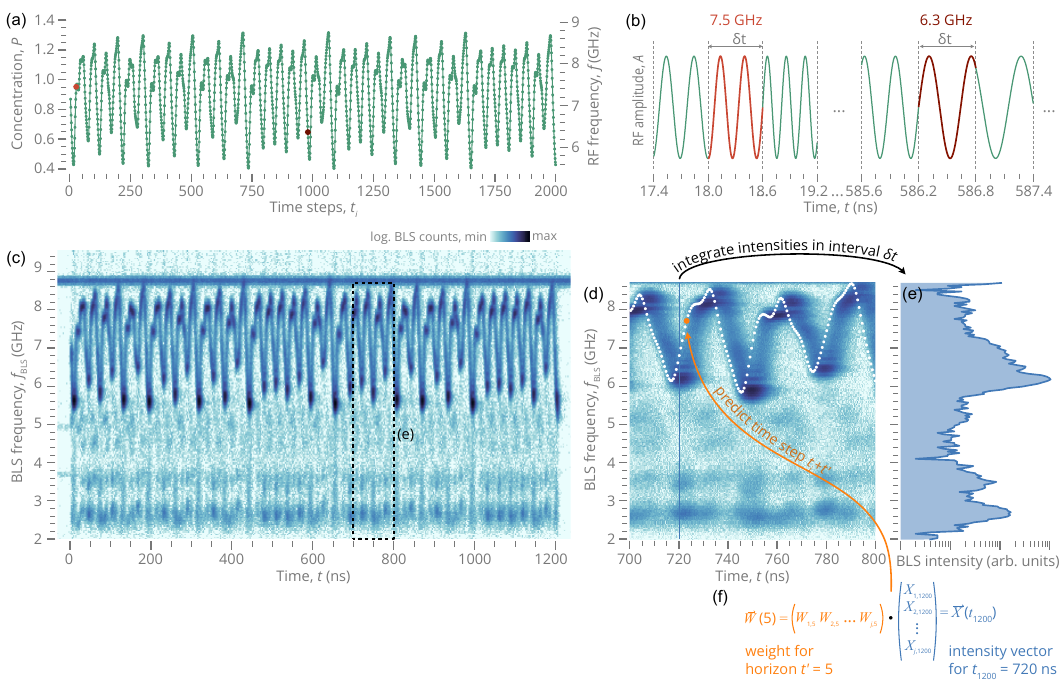}
    \caption{ (a) The Mackey-Glass time series is mapped onto a microwave current using continuous-phase frequency-shift keying. (b) Each microwave frequency is held for a certain time window of $\delta t = 0.6$\,\si{\nano\second} and injected into the $\Omega$-shaped antenna surrounding the MSR. (c) The time-resolved BLS spectrum reveals both direct responses and nonlinear magnon–magnon scattering, expanding the input into a richer spectral space. (d) Zoomed in section of the data from panel d. White dots indicate the input frequencies from the arbitrary waveform generator. The measured magnon intensities persist well beyond the direct input window, providing fading memory. (e) The integrated spectral intensities form state vectors that serve as inputs for a linear readout used to generate predictions for a given horizon $t^\prime$. (f) Definition of intensity vector and weight vector.}
   \label{fig:methodology}
\end{figure*}
%%%%%%%%%%%%%%%%%%%%%%%%%%%%

Here, we demonstrate chaotic time-series prediction using a magnon-scattering reservoir (MSR) operating with modal multiplexing~\cite{korber2023pattern,heinsBenchmarkingMagnonscatteringReservoir2025}. 
As shown in Fig.~\ref{fig:principle}, the platform consists of a Ni$_{81}$Fe$_{19}$ microdisk in the magnetic vortex state~\cite{cowburnSingleDomainCircularNanomagnets1999,scholzTransitionSingledomainVortex2003}, where magnetic moments curl in-plane around an out-of-plane vortex core. %This system provides key advantages essential for reservoir computing: strong intrinsic nonlinear magnon interactions, temporal memory arising from magnon dispersion and damping, and compatibility with complementary metal-oxide-semiconductor (CMOS) technology~\cite{heinsElectricalDetectionMagnons2025}. 
A time-series input, encoded in a microwave current via frequency modulation, is injected into a microwave antenna for exciting the magnetization inside the vortex-state disk. Strong intrinsic nonlinear scattering between various magnon modes project this one-dimensional, serial input signal into a higher-dimensional spectral output which is used to train a simple linear readout.
Our results show that the MSR reliably forecasts the dynamics described by the Mackey-Glass equation, even when only relying on a single unseen spectrum, and establish magnon-based reservoirs as a  physical computing platform for real-time prediction of complex dynamics.

%%%%%%%%%%%%%%%%%%%%%%%%%%%% 
\section{Methods}
%%%%%%%%%%%%%%%%%%%%%%%%%%%% 

%%%%%%%%%%%%%%%%%%%%%%%%%%%% 
\subsection{Time-resolved Brillouin light scattering microscopy}
%%%%%%%%%%%%%%%%%%%%%%%%%%%% 

To probe the MSR response, we use time-resolved micro-focused Brillouin light scattering (TR-µBLS) spectroscopy~\cite{sebastianMicrofocusedBrillouinLight2015}. Therefore, a continuous-wave laser (532\,nm) is focused onto the sample surface using a 100$\times$ objective lens (NA = 0.75), yielding a spatial resolution of approximately 300\,nm. The backscattered light is analyzed in a Tandem Fabry-Pérot interferometer (TFPI)~\cite{mockConstructionPerformanceBrillouin1987}, which resolves the frequency shift due to inelastic photon-magnon scattering with a spectral bin size of 37.5\,MHz. 

Photon counts, interferometer control signals, and a synchronized clock reference from the arbitrary waveform generator, which is used to excite magnons, are recorded by a time-to-digital converter (Timetagger 20, Swabian Instruments) with 200\,ps resolution. From these data, the temporal evolution of the magnon spectra with respect to the stroboscopic microwave excitation is reconstructed.

Sample drift is actively compensated using image recognition of in-situ CCD images of the structure, with corrections applied via high-precision stages (Newport XMS series). 
To access modes with different spatial profiles, µBLS spectra are averaged over 25 positions covering one quarter of the disk (five azimuthal angles at five radial positions). All measurements are performed at room temperature.

%%%%%%%%%%%%%%%%%%%%%%%%%%%% 
\subsection{Sample preparation and characterization}
%%%%%%%%%%%%%%%%%%%%%%%%%%%% 

The standard MSR consists of a 5\,\si{\micro\meter}-diameter magnetic disk patterned on a SiO$_2$ substrate, as shown in the scanning electron microscopy image in Fig.~\figref{fig:3ms}{a}. Fabrication employed a polymethyl methacrylate (950PMMA A4) resist mask, electron beam lithography, sputter deposition, and lift-off. The Ni$_{81}$Fe$_{19}$ (50\,nm)/Ta (4\,nm) bilayer was sputtered in a 1\,mT rotating magnetic field to suppress anisotropy ~\cite{lozhkinaTailoringMagneticProperties2023}. 

To excite magnetization dynamics, an $\Omega$-shaped microwave antenna was patterned around the disk using a double-layer resist stack (ethyl lactate EL11 and PMMA 950 A4), followed by electron beam lithography, deposition of Cr (5\,nm)/Au (150\,nm), and lift-off. The  inner and outer diameters of the antenna are 8.7 and 11\,\si{\micro\meter}, respectively. Microwave currents in the GHz range excite well-defined magnon eigenmodes of the vortex-state disk, which are characterized by their radial and azimuthal mode indices ($n,m$) counting the number of nodes along the disk radius and perimeter, respectively~\cite{buessExcitationsNegativeDispersion2005,vogtOpticalDetectionVortex2011}. 

Additional MSR geometries, including 3\,\si{\micro\meter}-diameter disks, rings (5\,\si{\micro\meter} outer diameter, hole widths of 525\,nm and 800\,nm) and squares (5\,\si{\micro\meter} edge length), were fabricated simultaneously on the same substrate and each embedded in an individual $\Omega$-shaped antenna [see Appendix~A, Fig.~\figref{suppfig:devices}{a-e}]. 

First, we focus on the 5\,\si{\micro\meter}-diameter disk. If radial modes therein are excited at microwave powers above threshold, nonlinear three-magnon splitting occurs~\cite{schultheiss_excitation_2019,verbaTheoryThreemagnonInteraction2021}. Figure~\figref{fig:3ms}{b} shows the mode amplitude profiles for an exemplary splitting process: the magnon mode $(2,0)$ initially excited at frequency $f_\text{initial}$ splits into two secondary magnons $(2,\pm 4)$ and $(0, \mp 4)$, under conservation of energy ($f_\text{initial} = f_{+} + f_{-}$) and angular momentum ($m_\text{initial} = m_{+} + m_{-}$).

The BLS spectra in Fig.~\figref{fig:3ms}{c} are measured for increasing excitation frequency and demonstrate that multiple of these three-magnon splitting channels exist. Excitation frequencies between 5.5 and 8.8\,\si{\giga\hertz} yield the strongest nonlinear response for the 5\,\si{\micro\meter}-diameter disk. 
As shown previously~\cite{korber2023pattern,heinsBenchmarkingMagnonscatteringReservoir2025}, the coexistence of multiple splitting channels leads to non-reciprocal mutual stimulation, such that different secondary modes are populated depending on the temporal order of various frequency componenents in a multitone excitation signal. This mechanism underpins the nonlinear expansion and temporal memory essential for the reservoir’s predictive capability.

%%%%%%%%%%%%%%%%%%%%%%%%%%%%
\begin{figure*}[]
    \centering
    \includegraphics{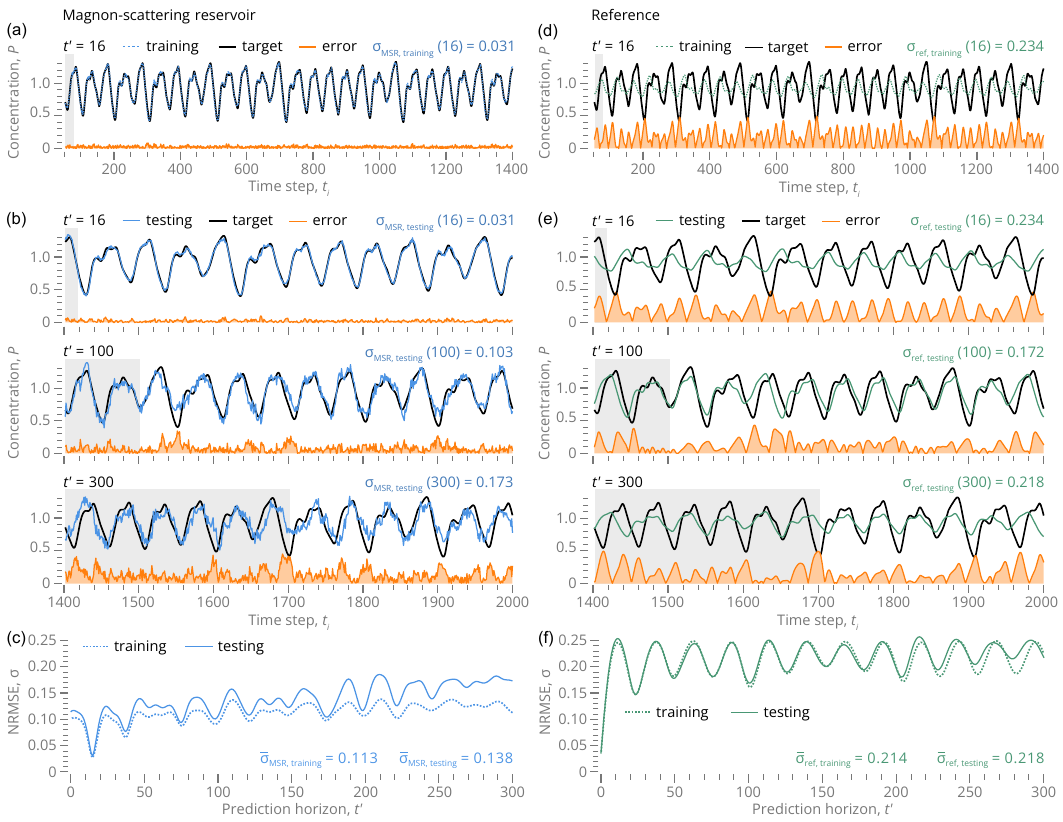}
    \caption{(a) Training data for the MSR when predicting $t^\prime = 16$ steps into the future, as indicated by the shaded area.  (b) Testing results for the MSR for increasing prediction horizon $t^\prime = 16$, 100, and 300 steps, as indicated by shaded areas. (c) Normalized root mean squared error (NRMSE) as a function of prediction horizon $t^\prime$, comparing training (dotted) and testing (solid). Viable forecasts are achieved even up to 300 steps, demonstrating long-range prediction performance. (d) Training data of the reference prediction task without MSR when predicting $t^\prime = 16$ steps into the future. (e) Testing results without the MSR for increasing prediction horizon $t^\prime = 16$, 100, and 300 steps. (f) NRMSE as a function of prediction horizon $t^\prime$, comparing training (dotted) and testing (solid) for the reference task. Note that in panels (a), (b), (d), and (e) the displayed errors correspond to the deviation of the training and testing outputs from the target MG data and are therefore larger than the NRMSE shown in the lower panels.}
    \label{fig:prediction}
\end{figure*}
%%%%%%%%%%%%%%%%%%%%%%%%%%%%

%%%%%%%%%%%%%%%%%%%%%%%%%%%% 
\subsection{Time-series generation and MSR response}
%%%%%%%%%%%%%%%%%%%%%%%%%%%% 

To evaluate our reservoir's prediction capabilities, we use the Mackey-Glass (MG) time series. This prototypical nonlinear model is capable of generating chaotic signals, which are widely employed to evaluate reservoir computing systems~\cite{mackey1977oscillation,RC_Benchmarks_York,gartsideReconfigurableTrainingReservoir2022}. The MG sequence is based on the biological process modelling the concentration $P(t)$ of mature red blood cells in living organisms mimicking the time delayed production of new cells in the bone marrow. Essentially, the MG sequence is generated from its delay-differential equation 
\begin{equation}
    \frac{dP(t)}{dt} = \frac{\beta P(t-\tau)}{1 + P(t-\tau)^{\alpha}} - \gamma P(t),
    \label{eq:MG}
\end{equation}
where $\tau$ is the delay time, $\beta$ the production rate, $\gamma$ the decay rate, and $\alpha$ the nonlinearity parameter. 

Depending on the choice of parameters the equation yields a periodic oscillation of the output value or a chaotic trajectory when the parameters are chosen for strong nonlinear dynamics. In Fig.~\figref{fig:methodology}{a}, we generated the MG sequence using the parameter set $\beta = 0.2$, $\gamma = 0.1$, $\alpha = 10$, and $\tau = 17$, which yields a chaotic trajectory. The equation was solved using the python package \textit{BrainPy} \cite{brainpy_mackeyglass}. From the resulting trajectory, we extracted 2000 discrete points $P(t_i)$, which served as the input sequence for the magnon reservoir.

To couple this signal into the MSR, we modulate a microwave current via continuous-phase frequency-shift keying. For each time step $t_i$ the value obtained from the MG equation is mapped onto a microwave signal within the frequency band yielding the strongest nonlinear response (5.5 to 8.8\,\si{\giga\hertz}), as shown by the right $y$-axis in Fig.~\figref{fig:methodology}{a}.

Each discrete frequency value was represented by a sinusoidal burst of duration $\delta t = 0.6$\,ns [Fig.~\figref{fig:methodology}{b}],  synthesized by an arbitrary waveform generator (\textit{AWG7000 Tektronix}) at a sampling rate of \SI{25}{GS/s} and $V_{\mathrm{pp}}$ of \SI{500}{mV}. To ensure a continuous waveform, the relative phase between consecutive bursts was adjusted during synthesis.
This signal is transmitted through a series of amplifier (\SI{+23}{dBm}), attenuator (\SI{-5}{dBm}), and amplifier (\SI{+16}{dBm}). Additionally, a signal generator (\textit{Keysight N5173B}) provides a continuous-wave \SI{8.8}{GHz} microwave signal at \SI{25}{dBm}, above the nonlinear threshold, to increase the number of active magnon modes.
Both signals are merged using a frequency combiner and applied to the $\Omega$-shaped antenna surrounding the MSR via a ground-signal-ground microwave probe. The effect of choosing different time windows is discussed in Appendix~B. 

Figure~\figref{fig:methodology}{c} shows the spectral intensity measured with time-resolved BLS during the MG sequence. Within the direct excitation band, the response follows the input frequencies. More importantly, strong signals also appear in the frequency range from 2 to 5.5\,GHz due to nonlinear three-magnon splitting. This intrinsic nonlinear mode coupling is essential for the MSR's performance. The nonlinearity not only depends on the system's history and provides a unique fingerprint of the dynamic state in which the system is evolving, it also projects the one-dimensional input into a much higher-dimensional spectral output.

In addition, the zoomed in spectrum in Fig.~\figref{fig:methodology}{d} highlights that the magnon responses to the microwave excitation signal persist well beyond the 0.6\,\si{\nano\second} input window, contributing to the fading memory. These two features – nonlinear expansion and memory – are the fundamental ingredients for reservoir computing.

To define the reservoir states, the BLS intensity is integrated within each 0.6\,ns input window across the entire frequency range from 2 to 8.8\,\si{\giga\hertz}, covering the directly excited modes as well as the nonlinear response, as displayed in Fig.~\figref{fig:methodology}{e}. The resulting intensities are assembled into a spectral vector $\vec{X}(t_i)$ for each time step [Fig.~\figref{fig:methodology}{f}]. The number of entries $j=182$ of these intensity vectors is determined by the frequency bin size in the BLS measurement, which is set to 37.5\,\si{\mega\hertz}. Concatenation of all vectors yields the reservoir state matrix $\mathbf{X}$, which represents the nonlinear transformation of the MG input by our MSR.

%%%%%%%%%%%%%%%%%%%%%%%%%%%%
\subsection{Training procedure and error metrics}
%%%%%%%%%%%%%%%%%%%%%%%%%%%%

The prediction task was formulated as learning a linear mapping between reservoir states and MG values at a future time step $t^\prime$:  
\begin{equation}
    \vec{y}_\mathrm{training}(t^\prime) = \vec{W}_{t^\prime}\cdot \mathbf{X} + \mathrm{const.}
\end{equation}
with $\vec{y}_\text{training}(t^\prime) = \big(P(51+t^\prime), P(52+t^\prime), ..., P(1400+t^\prime) \big) $ and $\vec{W}_{t^\prime}$ the weight vector trained for each prediction horizon $t^\prime$. Training and testing sets were created by splitting the MG sequence into non-overlapping subsets of 1350 ($51 \leq t_i \leq 1400$) and 600 ($1401 \leq t_i \leq 2000$) time steps, discarding the first 50 steps to allow the system to reach steady operation. 

The weight vector and constant were found using the \textit{LinearRegression} model from \textit{scikit-learn} \cite{scikit-learn}, where the model only uses an analytical method to find the suitable parameters without any iterative optimization. This also means that the training time is very short.

To put the MSR's performance into perspective, we follow the same training routine directly on the MG data $P(t_i+t^\prime)=w_{t^\prime} P(t_i)+\text{const}.$ with weights $w_{t^\prime}$, assuming a scenario without the MSR projecting the time-series input to the higher-dimensional spectral output space. For this reference task, we use the same learning algorithms as described above. 

The prediction performance is quantified by calculating the root mean squared error (RMSE) between the predicted $\hat{y}_i(t^\prime)$ and target $y_i(t^\prime)$ values:  
\begin{equation}
    \mathrm{RMSE} = \sqrt{\frac{1}{n}\sum_{i=1}^n (y_i(t^\prime) - \hat{y}_i(t^\prime))^2}.
\end{equation} 
To allow comparison across different signal amplitudes, the RMSE is normalized to the dynamic range of the MG sequence, yielding the normalized root mean squared error (NRMSE):  
\begin{equation}
    \mathrm{NRMSE} = \sigma(t^\prime) = \frac{\mathrm{RMSE}}{y_{\max} - y_{\min}}.
    \label{eq:NRMSE}
\end{equation} 

A smaller NRMSE indicates higher predictive accuracy. For each prediction horizon $t^\prime$, the error is computed separately for the training and testing sets to monitor potential overfitting.  In addition, the mean prediction error $\bar{\sigma}$ is defined as the average NRMSE over all tested horizons from $t^\prime=1$ to 300 steps  
\begin{equation}
    \bar{\sigma} = \frac{1}{300} \sum_{t^\prime=1}^{300} \sigma(t^\prime),
\end{equation}
providing a single figure of merit for overall forecasting performance.

%%%%%%%%%%%%%%%%%%%%%%%%%%%% 
\begin{figure}[]
    \centering
    \includegraphics{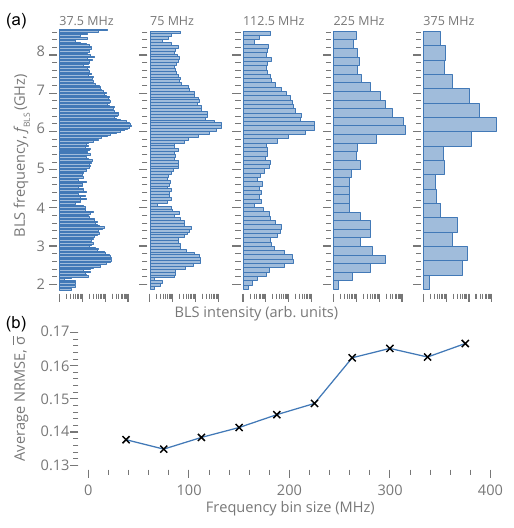}
\caption{(a) BLS spectra extracted with different frequency bin sizes. (b) Average NRMSE as a function of bin size, showing an optimal performance at twice the natural BLS spectral resolution. Excessive binning reduces spectral richness and degrades forecasting accuracy.}
    \label{fig:binning}
    \end{figure}
%%%%%%%%%%%%%%%%%%%%%%%%%%%% 

%%%%%%%%%%%%%%%%%%%%%%%%%%%% 
\begin{figure}[]
    \centering
    \includegraphics{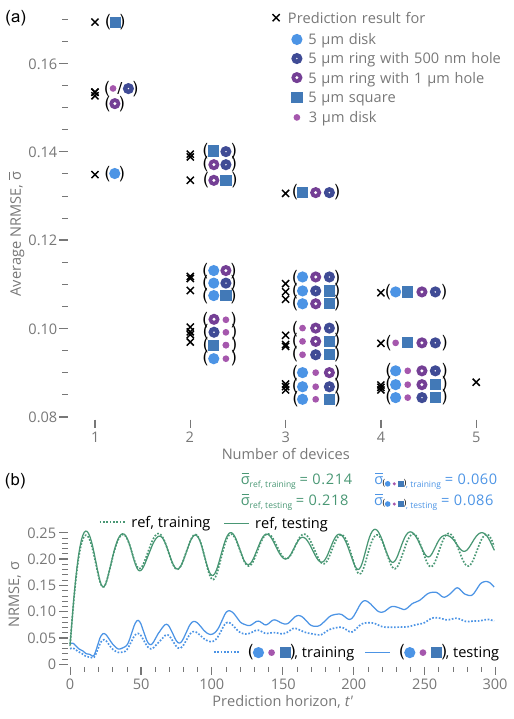}
    \caption{(a) Average NRMSE for different MSR geometries and their combinations. Appending outputs from multiple devices systematically reduces error, showing that greater reservoir depth enhances forecasting performance. (b) NRMSE as a function of prediction horizon $t^\prime$, comparing training (dotted) and testing (solid) for the optimum combination of devices and the reference task.}
    \label{fig:devices}
\end{figure}
%%%%%%%%%%%%%%%%%%%%%%%%%%%% 

%%%%%%%%%%%%%%%%%%%%%%%%%%%% 
\section{Results}
%%%%%%%%%%%%%%%%%%%%%%%%%%%% 

For efficient forecasting, it is essential to monitor efficient training, as exemplary shown in  Fig.~\figref{fig:prediction}{a} for a horizon $t^\prime = 16$. Figure~\figref{fig:prediction}{b} shows successful forecasts for $t^\prime = 16$, 100, and 300 steps, with errors increasing gradually with horizon length. Determining an average NRMSE for each prediction horizon $t^\prime$, as plotted in Fig.~\figref{fig:prediction}{c}, shows that the error remains low, even when the MSR predicts 300 steps into the future. 
The results for the reference task are summarized in Fig.~\figref{fig:prediction}{d-f}. Overall, the  errors remain well above the ones achieved with the MSR across all prediction horizon.

As mentioned before, the dimensionality of the reservoir state is determined by the frequency bin size of the BLS output.
Allowing systematic evaluation of the trade-off between feature richness and regression complexity, we applied additional binning to the spectra and, thereby, to the intensity state vectors, effectively reducing the number of features per time step as shown in Fig.~\figref{fig:binning}{a}. Within each frequency bin, the detected BLS intensity is averaged. Smaller state dimensions simplify the regression task by lowering the number of trainable parameters, which can improve stability during training and mitigate overfitting for limited datasets. This can be seen in Fig.~\figref{fig:binning}{b} when doubling the bin size from the BLS inherent 37.5\,\si{\mega\hertz} bins to 75\,\si{\mega\hertz}. 

However, excessive binning discards spectral detail, thereby reducing the richness of the nonlinear mapping and degrading prediction performance. We find that moderate binning yields a favorable compromise between computational efficiency and accuracy, whereas large bin sizes significantly diminish the forecasting horizon. This highlights the importance of tuning the reservoir dimensionality to balance training efficiency and predictive power.

Thus far, our predictions were obtained from a single 50\,\si{\nano\meter}-thick and 5\,\si{\micro\meter}-wide vortex-state MSR. To explore how reservoir richness influences performance, we extend the approach including four other distinct device geometries, all excited via $\Omega$-shaped antennas: a 3\,\si{\micro\meter}-diameter disk, two rings of 5\,\si{\micro\meter} outer diameter with hole widths of 525\,nm and 800\,nm, and a 5\,\si{\micro\meter} square. On the one hand, additional geometries were chosen similar to the previously studied 5\,\si{\micro\meter} disk to yield adequate nonlinear response when excited with microwave currents in the same frequency range. On the other hand, adding smaller disks and other shapes increased diversity. However, we want to point out that the choice of additional geometries was not optimized yet but already demonstrates the feasibility of increasing the reservoir complexity.

The input waveform was identical across devices, designed based on the spectral properties of the 5\,\si{\micro\meter} disk, and the outputs were measured using the same BLS setup (see Appendix~A, Fig.~\ref{suppfig:devices} for the temporal responses of the other devices).

To construct an enlarged reservoir state, the frequency-intensity vectors (with 75-MHz binning) from different geometries were combined, forming larger state vectors $\vec{X}(t_i)$. The same training and testing protocol was applied, enabling direct comparison of predictive performance across different reservoir depths. As summarized in Fig.~\figref{fig:devices}{a}, prediction performance improves with the number of devices included. In particular, the combination of the 3\,\si{\micro\meter} and 5\,\si{\micro\meter} disks with the 5\,\si{\micro\meter} square yields the lowest average NRMSE across the entire prediction horizon, as highlighted in Fig.~\figref{fig:devices}{b}.  

%%%%%%%%%%%%%%%%%%%%%%%%%%%% 
\begin{figure}[]
    \centering
    \includegraphics{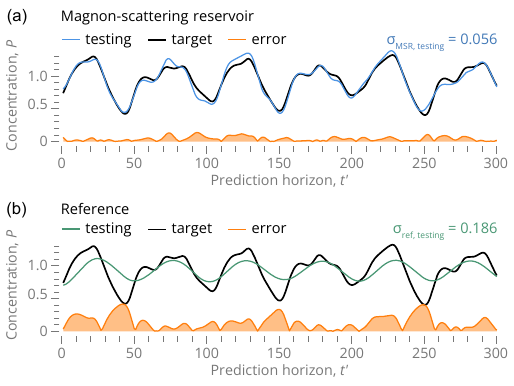}
    \caption{(a) Predicting 300 steps of the MG series directly from one unseen spectrum of the MSR by concatenating 300 individually trained prediction models into a single weight matrix. This approach yields consistently low errors across all horizons. (b) In stark contrast, models trained only on MG data show much lower accuracy.}
    \label{fig:300models_pred}
\end{figure}
%%%%%%%%%%%%%%%%%%%%%%%%%%%% 

Ultimately, we concatenate the 300 individually trained prediction models $\vec{W}_{t^\prime}$ for the optimal device combination (3\,\si{\micro\meter} and 5\,\si{\micro\meter} disks with the 5\,\si{\micro\meter} square) into a single weight matrix $\mathbf{W}$. This allows forecasting up to 300 future steps directly from one unseen spectrum $\vec{X}(t_i=1401)$:
\begin{equation}
    \vec{y}_\text{testing} = \mathbf{W}\cdot \vec{X}(1401) + \text{const.}
    \label{eq:LR}
\end{equation}
with $\vec{y}_\text{testing} = \big( P(1402), P(1403), ..., P(1701) \big) $.
As shown in Fig.~\figref{fig:300models_pred}{a}, this approach yields consistently low errors, much lower compared to the reference prediction model without the magnon reservoir plotted in Fig.~\figref{fig:300models_pred}{b}.

Beyond enhancing accuracy, this approach also changes the prediction paradigm. Instead of relying on long sequences of MG input data, the concatenated models  extrapolate far into the future from a single reservoir state.
This ability to forecast hundreds of steps from a single reservoir state underscores the richness of the magnonic representation and illustrates a fundamentally new prediction paradigm, where the future trajectory can be reconstructed from a single snapshot rather than a continuous input stream.

%%%%%%%%%%%%%%%%%%%%%%%%%%%% 
\section{Discussion and Outlook}
%%%%%%%%%%%%%%%%%%%%%%%%%%%% 

Our results demonstrate that a magnon-scattering reservoir can reliably forecast the Mackey-Glass time series over hundreds of time steps, establishing magnons as a viable substrate for physical reservoir computing. A prediction horizon of $t^\prime = 300$ corresponds to approximately six full oscillation cycles of the Mackey-Glass sequence, indicating that the magnon reservoir can reliably capture long-term variations, substantially exceeding previously demonstrated magnetic reservoirs based on artificial spin-ice arrays~\cite{gartsideReconfigurableTrainingReservoir2022,stenningNeuromorphicOverparameterisationFewshot2024,leeTaskadaptivePhysicalReservoir2024}, which achieved short-term prediction over one to two characteristic cycles with comparable errors. This improvement arises from the continuous, GHz-frequency dynamics of magnons, which provide both stronger nonlinearity and intrinsic fading memory on nanosecond timescales.

Beyond benchmarking on the Mackey-Glass system, the intrinsic properties of magnons suggest several directions for future exploration. First, strong nonlinear interactions and long-lived excitations provide a natural resource for tackling tasks that require both high-dimensional feature spaces and fading memory, such as speech recognition or real-time sensor data processing. Second, the frequency-tunable nature of magnons opens pathways toward multiplexing and parallelism, potentially enabling multi-channel prediction on the same chip. Third, compatibility with CMOS processes points toward scalable integration, where arrays of magnetic microdisks could be co-fabricated with conventional electronics.

An additional design principle emerges from our analysis of spectral binning. While reducing dimensionality can simplify training and lower computational cost, excessive compression comes at the expense of predictive accuracy. This trade-off underscores the importance of tailoring the effective state dimension of the reservoir to the specific task at hand. In scalable architectures, such control could be realized by dynamically adjusting spectral resolution or by combining multiple reservoirs with complementary dimensionalities. More broadly, this highlights that not only the physical substrate but also the way its output is encoded plays a crucial role in optimizing performance.

From a broader perspective, our work connects nonlinear magnetism with machine learning and neuromorphic computing, demonstrating how fundamental spin-wave physics can be leveraged for tasks traditionally reserved for artificial neural networks. 
The demonstrated ability to map a simple one-dimensional input into a rich spectral output illustrates how magnonic devices can function as compact, intrinsically energy-efficient reservoirs, where nonlinear transformation and memory emerge naturally from low-loss spin dynamics instead of being emulated through large numbers of active electronic elements.
Looking ahead, combining magnon reservoirs with other spintronic or photonic elements may yield hybrid architectures that exploit the strengths of multiple substrates. Such advances could bring physical reservoir computing closer to deployment in edge devices, where real-time prediction and low power consumption are critical.

In summary, the use of magnons for chaotic time-series prediction highlights the potential of magnonics as a platform for unconventional computing. By uniting the fields of nonlinear dynamics, magnetism, and machine learning, this approach opens new avenues for scalable, real-time information processing in both scientific and technological contexts.

%%%%%%%%%%%%%%%%%%%%%%%%%%%% 
\section*{Acknowledgements}
%%%%%%%%%%%%%%%%%%%%%%%%%%%% 

The project has received funding by the EU Research and Innovation Programme Horizon Europe under grant agreement no. 101070290 (NIMFEIA). Support by the Nanofabrication Facilities Rossendorf (NanoFaRo) at the IBC is gratefully acknowledged.

H.S. and K.S. conceptualized the presented work.
M.K., J.F., H.S., and K.S. acquired funding.
F.K. and K.S. fabricated the sample.
Z.X., C.H., and T.D. conducted the experiments. Z.X. and C.H. analyzed the data.
%performed and analyzed the micromagnetic simulations.
Z.X., H.S., and K.S. visualized the results.
Z.X. and K.S. wrote the original draft of the paper.
All authors reviewed and edited the paper.

The authors have no conflicts of interest to disclose.

%%%%%%%%%%%%%%%%%%%%%%%%%%%% 
\section*{Data availability}
%%%%%%%%%%%%%%%%%%%%%%%%%%%% 
The data that support the findings of this study are openly available in Ref.~\cite{xiong_zeling_2025_4028}.

%%%%%%%%%%%%%%%%%%%%%%%%%%%% 
\section*{Appendix A: Time-resolved BLS spectra for different geometries}
%%%%%%%%%%%%%%%%%%%%%%%%%%%% 

\begin{figure*}[]
    \centering
    \includegraphics{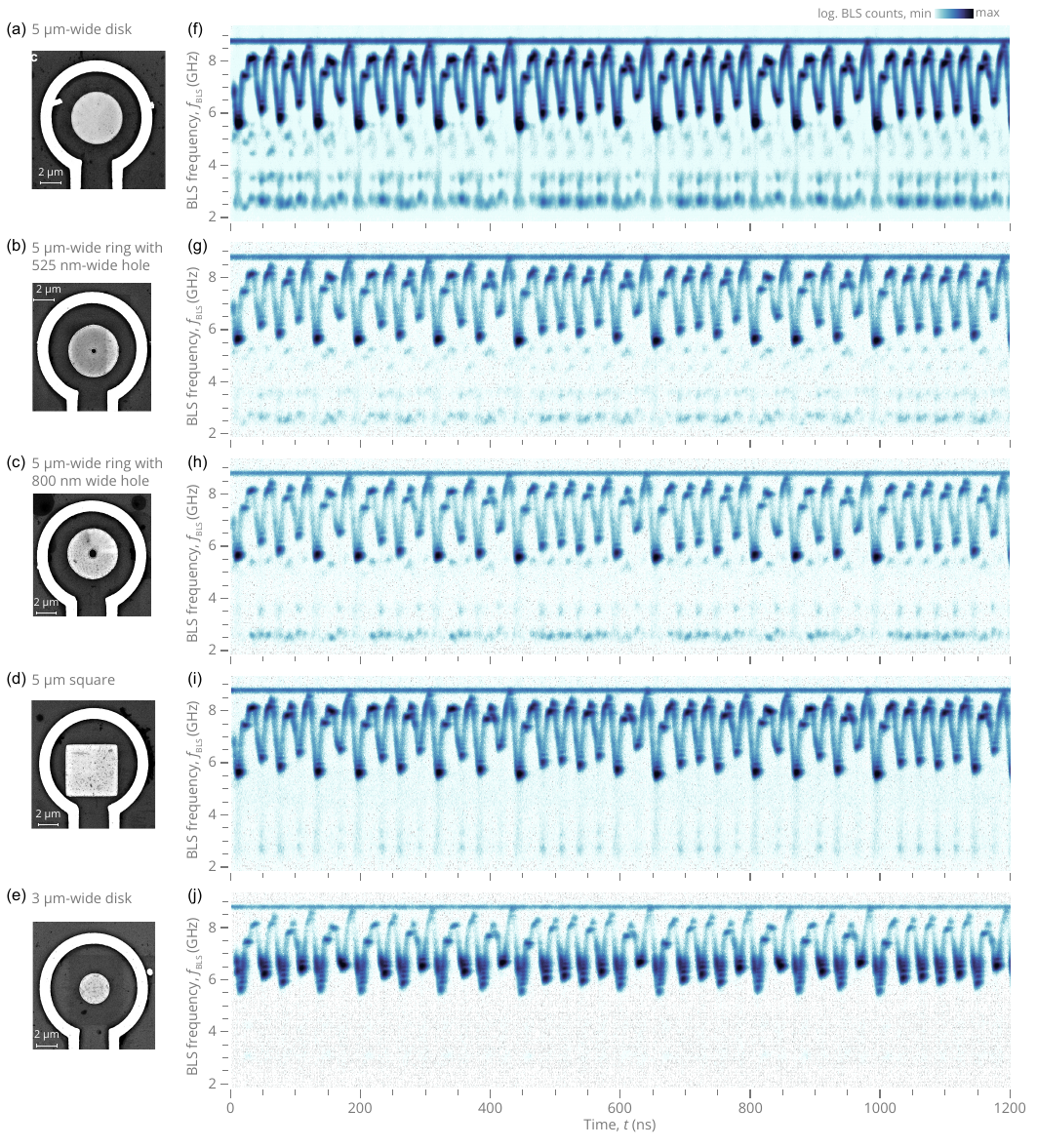}
    \caption{ (a)-(e) Scanning electron microscopy images of magnon reservoirs with different geometries. (f)-(j) Time-resolved BLS spectra measuring the different reservoirs' response to the complex MG time-series sequence. The nonlinear response recorded for BLS frequencies between 2 and 5.5\,\si{\giga\hertz} gets weaker the more the geometry diverts from the 5\,\si{\micro\meter}-diameter disk. This is to be expected since the frequency range for the microwave current inputting the MG sequence was optimized to address the fundamental modes in the \SI{5}{\micro\meter}-diameter disk. }
    \label{suppfig:devices}
\end{figure*}

For increasing the reservoir dimensionality, we input the same frequency-modulated MG data into five different devices: two disks of  5\,\si{\micro\meter} and 3\,\si{\micro\meter} diameter disk, two rings of 5\,\si{\micro\meter} outer diameter with hole widths of 525\,nm and 800\,nm, and a 5\,\si{\micro\meter} square. Scanning electron microscopy images of all devices are shown in Fig.~\figref{suppfig:devices}{a-e}. 

On each individual device we recorded time-resolved BLS spectra, integrated over multiple scan positions. The spectra are plotted in Fig.~\figref{suppfig:devices}{f-j}. Since the frequency range for the MG input was designed to yield the maximum nonlinear response for the 5\,\si{\micro\meter} disk, the spectra for other devices show less nonlinearity, the more their geometry diverges from the 5\,\si{\micro\meter} disk. This also explains why the 5\,\si{\micro\meter} disk gives the best accuracy among all individual predictions [Fig.~\figref{fig:devices}{a}].

%%%%%%%%%%%%%%%%%%%%%%%%%%%% 
\section*{Appendix B: Influence of input duration}
%%%%%%%%%%%%%%%%%%%%%%%%%%%%

\begin{figure*}[]
    \centering
    \includegraphics{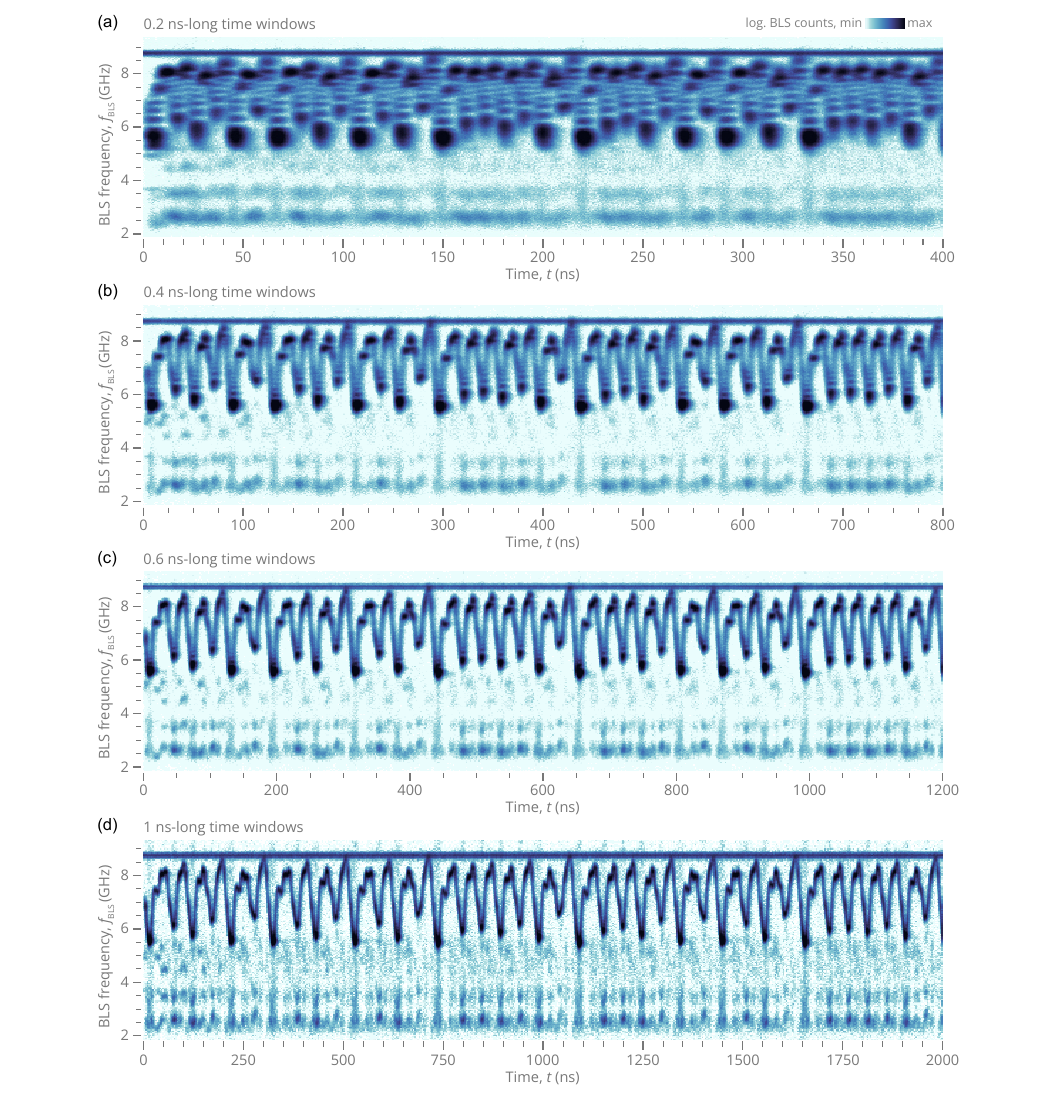}
    \caption{Time-resolved BLS spectra measured for the MG sequence encoded using different time windows $\delta t$ of (a) 0.2\,\si{\nano\second}, (b) 0.4\,\si{\nano\second}, (c) 0.6\,\si{\nano\second}, and (d) 1\,\si{\nano\second}. The 0.2\,\si{\nano\second}-long input is too fast to allow the magnon system to evolve a clear dynamic response. With increasing $\delta t$, more nonlinear scattering sets in. However, if signals are too long, they have less overlap with past data due to the limited lifetime of magnons. Thus, the balance between system response and enough data overlap has to be found. By comparing their average NRMSE values without binning, we determined $\delta t = 0.6$\,\si{\nano\second} to be the best duration for inputting the microwave current via continuous-phase frequency-shift keying. }
    \label{suppfig:times}
\end{figure*}

The MSR's response to the MG sequence clearly depends on the time scale with which the frequency-modulated microwave current is applied. Therefore, we test different signal duration $\delta t$ between \SIrange{0.2}{1}{\nano\second} to find the time scale optimal for future prediction. The resulting time-resolved BLS spectra are summarized in Fig.~\ref{suppfig:times}. By comparing NRMSE values for different time windows, best results are obtained for  \SI{0.6}{\nano\second}. This relates to the balance between reaction time, given by the onset of nonlinearity, and memory capacity in our system, mostly determined by the intrinsic magnon lifetime.

%%%%%%%%%%%%%%%%%%%%%%%%%%%% 
%%% References
%%%%%%%%%%%%%%%%%%%%%%%%%%%% 
\bibliographystyle{apsrev4-2-titles}
\bibliography{references.bib}
%%%%%%%%%%%%%%%%%%%%%%%%%%%% 

%%%%%%%%%%%%%%%%%%%%%%%%%%%% 
%%%%%%%%%%%%%%%%%%%%%%%%%%%% 

%%%%%%%%%%%%%%%%%%%%%%%%%%%% 
%%%%%%%%%%%%%%%%%%%%%%%%%%%% 

\end{document}